\documentclass[letter]{aa}         

\usepackage{natbib}
\usepackage{graphicx}
\usepackage{txfonts}
\usepackage{hyperref}

\newcommand{\planck}{\textsl{Planck}}
\newcommand{\asca}{\textsl{ASCA}}
\newcommand{\rosat}{\textsl{ROSAT}}
\newcommand{\chandra}{\textsl{Chandra}}
\newcommand{\xmm}{\textsl{XMM-Newton}}
\newcommand{\suzaku}{\textsl{SUZAKU}}
\newcommand{\erosita}{\textsl{SRG/eROSITA}} 

\newcommand{\beq}{\begin{equation}}
\newcommand{\eeq}{\end{equation}}
\newcommand{\beqa}{\begin{eqnarray}}
\newcommand{\eeqa}{\end{eqnarray}}
\newcommand{\mpc}{$h^{-1} \mathrm{Mpc}$}

\newcommand{\ie}{i.e.\xspace}

\def\msun{{\rm M}_{\odot}}

\def\zsun{Z_{\odot}}

\begin{document} 

  \title{First detection of stacked X-ray emission \\ from cosmic web filaments}
  \author{H. Tanimura \and N. Aghanim \and A. Kolodzig \and M. Douspis \and N. Malavasi 
  }

  \institute{
  Universit\'{e} Paris-Saclay, CNRS, Institut d'Astrophysique Spatiale, B\^atiment 121, 91405 Orsay (France) \\
  \email{hideki.tanimura@ias.u-psud.fr}
             }
             
  \date{}

  \abstract 
    {We report the first statistical detection of X-ray emission from cosmic web filaments in \rosat{} data. We selected 15,165 filaments at 0.2<$z$<0.6 ranging from 30 Mpc to 100 Mpc in length, identified in the Sloan Digital Sky Survey (SDSS) survey. We stacked the X-ray count-rate maps from \rosat{} around the filaments, excluding resolved galaxy groups and clusters above the mass of $\sim3 \times 10^{13} \, \msun$ as well as the detected X-ray point sources from the \rosat, \chandra,\ and \xmm\ observations. The stacked signal results in the detection of the X-ray emission from the cosmic filaments at a significance of 4.2$\sigma$ in the energy band of 0.56--1.21 keV. The signal is interpreted, assuming the Astrophysical Plasma Emission Code (APEC) model, as an emission from the hot gas in the filament-core regions with an average gas temperature of $0.9^{+1.0}_{-0.6}$ keV and a gas overdensity of $\delta \sim 30$ at the center of the filaments. 
    Furthermore, we show that stacking the \erosita\ data for $\sim$2,000 filaments only would lead to a $\gtrsim5\sigma$ detection of their X-ray signal, even with an average gas temperature as low as $\sim$0.3 keV.}

\maketitle


\section{Introduction}
\label{sec:intro}

The cosmic structure is organized in a complex web-like pattern called cosmic web \citep{Bond1996} made of nodes, filaments, sheets, and voids (\ie, \citealt{Aragon2010, Cautun2013}). Numerical simulations predict that the majority of baryons are found in filaments in the form of hot plasma \citep{Cen1999, cen2006, Aragon2010, Cautun2014, Martizzi2019}: a temperature range of $10^5$--$10^7$ K and densities of $\sim$10--100 times the average cosmic value, called the warm hot intergalactic medium (WHIM). Extensive searches for the WHIM were performed in far-ultraviolet, X-ray, and thermal Sunyaev-Zel'dovich (tSZ) effect \citep{Zeldovich1969, Sunyaev1970, Sunyaev1972}. However, the measurement is difficult because the signal is relatively weak and the morphology of the source is complex. Some detections of gas in filaments have been reported with X-ray and tSZ observations. But these have been reported mostly in short and dense filaments, for example, between cluster pairs \citep{Fujita2008, Planck2013IR-VIII, Bonjean2018, Dietrich2005, Werner2008, Tittley2001, Alvarez2018, Sugawara2017} or extending beyond the virial radius of galaxy clusters \citep{Vikhlinin2013, Eckert2015, Connor2018, Connor2019}.
Gas residing in filaments can also be studied through absorption lines in the spectrum of active galactic nuclei (AGN) \citep{Nicastro2018, Kovacs2019}. However this has so far been limited to a few lines of sight where very X-ray bright AGNs are aligned with filaments.

Statistical detections of gas in stacked short filaments, on the order of 10 \mpc\ , have also been reported with tSZ observations \citep{Tanimura2019b, deGraaff2019}. Longer cosmic filaments on scales of 30--100 Mpc were studied in (\citealt{Tanimura2020}, hereafter T20) using the catalog of filaments detected in the Sloan Digital Sky Survey (SDSS) survey by \cite{Malavasi2020arXiv} and hot gas was detected by stacking tSZ measurements. However, additional probes are necessary to break the degeneracy between the gas density and temperature. In this paper, we study the filaments with the \rosat\ maps, which can be used to break the degeneracy and estimate the gas density and temperature. The paper is organized as follows: We describe data in Section \ref{sec:data} and methodology in Section \ref{sec:ana}, followed by results. In Section \ref{sec:erosita}, we assess a detectability of X-ray emission from gas in filaments by \erosita \citep[extended ROentgen Survey with an Imaging Telescope Array,][]{predehl2010}. We describe possible systematics in Section \ref{sec:systematics} and end with discussion and conclusions in Section \ref{sec:discussion} and Section \ref{sec:conclusion}. 

Throughout this work, we adopt the $\Lambda$CDM cosmology from \cite{planck2016-xiii} with $\Omega_{\rm m} = 0.3075$, $\Omega_{\Lambda} = 0.6910$, and $H_0 = 67.74$ km s$^{-1}$ Mpc$^{-1}$. The quantity $M_{\Delta}$ is the mass enclosed within a sphere of radius $R_{\Delta}$ such that the enclosed density is $\Delta$ times the critical density at redshift $z$. Uncertainties are given as 1$\sigma$ confidence level.


\section{Data}
\label{sec:data}

Following T20, we studied the filaments identified in \cite{Malavasi2020arXiv} using the Discrete Persistent Structure Extractor (DisPerSE) algorithm \citep{Sousbie2011}. This method computes the gradient of a density field and, where the gradient is zero, identifies critical points (maxima, minima, saddles, and bifurcations).\ The DisPerSE algorithm then defines filaments as field lines of constant gradient that connect maxima and saddles. The critical points are possible locations of galaxy groups and clusters and, for our purpose of studying filaments, the critical points and galaxy clusters were masked in our analysis as in T20. In the present analysis, we studied 15,165 filaments out of the 24,544 used in T20, the discarded filaments being covered by the larger mask used in this work (see Sect. \ref{subsec:stacking}).

We used the publicly available \rosat\ soft X-ray diffuse background (SXRB) maps \cite{snowden1997} covering almost the entire sky. These maps consist of 72 fits files, containing count rates and errors, derived from the 1990–1991 \rosat\ survey observations. Additional data were collected in 1997 to fill in the missing parts \citep{Freyberg1999}. These X-ray maps in HEALpix format are provided\footnote{http://www.jb.man.ac.uk/research/cosmos/rosat/} in six standard energy bands covering 0.1–2 keV (R1:0.11-0.284 keV, R2:0.14–0.284 keV, R4:0.44-1.01 keV, R5: 0.56-1.21 keV, R6:0.73-1.56 keV, and R7: 1.05-2.04 keV). We did not use the R3 band because the band has a significant response on both sides of the carbon absorption edge of the position sensitive proportional counter (PSPC) window and is in general not useful, as seen in Fig.1 in \cite{Snowden1994}

In this study, we also masked detected X-ray point sources. This includes 135,118 point-like sources in the Second ROSAT All-Sky Survey Point Source Catalog (2RXS)\footnote{http://www.mpe.mpg.de/ROSAT/2RXS}\citep{boller2016}, 317,167 X-ray sources in the Chandra Source Catalog (CSC 2.0)\footnote{http://cxc.cfa.harvard.edu/csc2/}\citep{evans2010}, and 775,153 sources in the third generation catalog of X-ray sources from \xmm\ observatory (3XMM-DR8)\footnote{https://www.cosmos.esa.int/web/xmm-newton/xsa\#download}\citep{rosen2016}.


\section{Data analysis}
\label{sec:ana}

\subsection{Measuring the X-ray profiles of the filaments}
\label{subsec:stacking}

    \begin{figure}
    \centering
    \includegraphics[width=\linewidth]{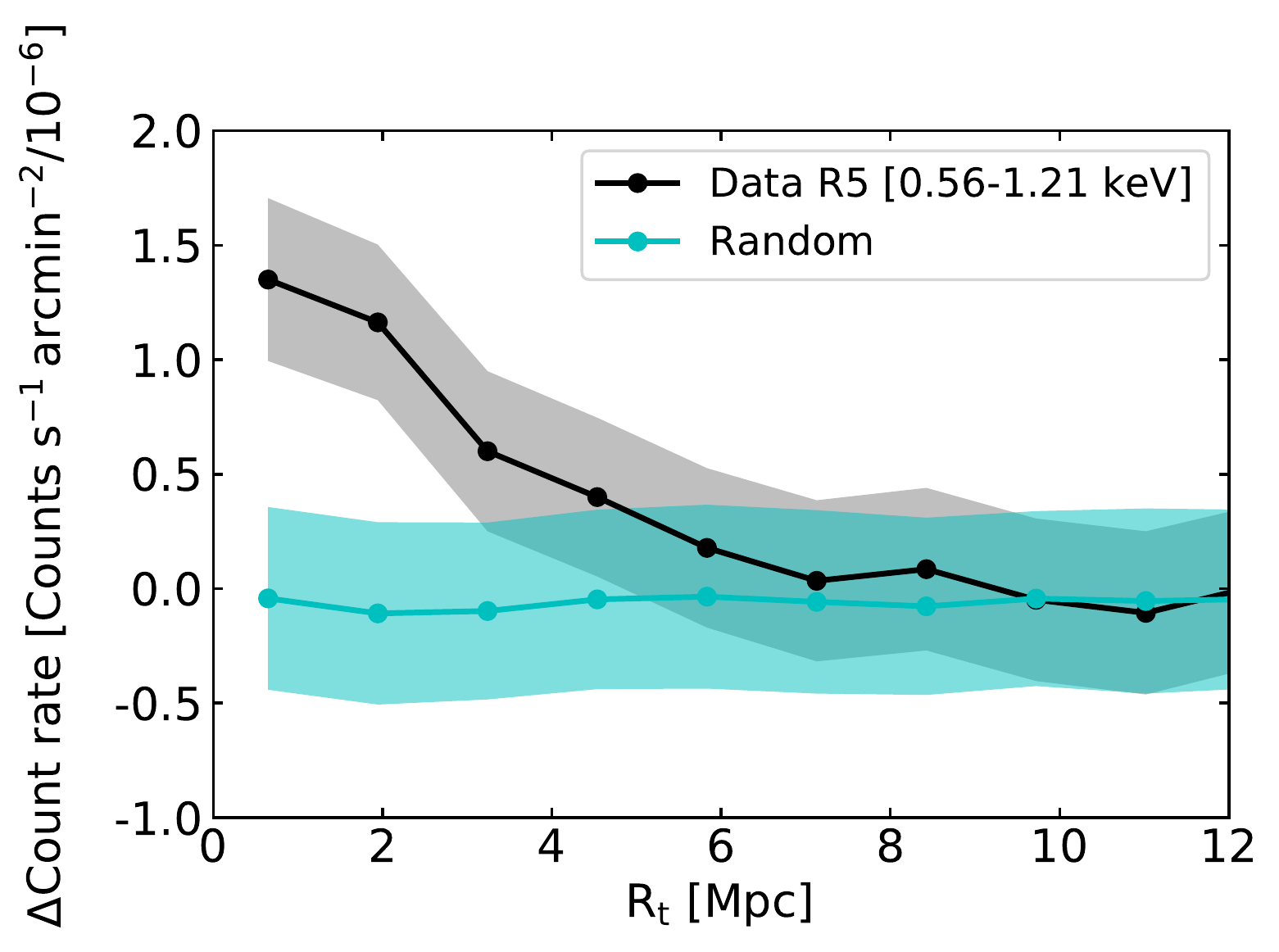} 
    \caption{Black: Average radial X-ray profile of 15,165 filaments in the \rosat\ R5 band. The 1$\sigma$ uncertainties estimated by a bootstrap resampling are shown in gray. Cyan: Average radial X-ray profile from 1,000 random sampling. For each random sampling, 15,165 filaments are displaced at random positions on the \rosat\ map. The 1$\sigma$ uncertainties are estimated by computing a standard deviation of the 1,000 random sample. }
    \label{fig:xprof}
    \end{figure}
    
We measured X-ray profiles of the 15,165 filaments using the \rosat\ maps at the filament positions. To remove the contamination from galaxy groups and clusters, the galaxy clusters were masked with a radius of $3 \times R_{500}$, and critical points (maxima and bifurcation with overdensities larger than 5) were masked with a radius of 10 arcmin, as in T20. Other critical points are underdense regions and not masked. 
In addition for the present study, which focuses on diffuse X-ray emission, we masked the X-ray point sources with a radius of 2 arcmin (on the order of the resolution of \rosat\ PSPC camera and significantly larger than those of \chandra\ and \xmm{}). In Sect. \ref{sec:systematics}, we investigate possible biases due to galaxy groups and clusters as well as X-ray point sources; we find these biases to be minor.
After this step, for each filament we extracted the region within 20 Mpc from a filament spine on the X-ray maps and computed a radial profile of the filament. We subtracted from the profile the average value within 10--20 Mpc from the filament spine, considered as a local background signal, which includes the Galactic and extragalactic cosmic X-ray background (CXB) emission. 
After repeating this step for all the filaments, we obtained a collection of 15,165 ``background-subtracted'' radial profiles and their average profile is shown in Fig.~\ref{fig:xprof}. (see (i)-(iv) in Sect. 3.1 of T20 for details.)  We assessed uncertainties by bootstrap resampling ({gray} in Fig.~\ref{fig:xprof}) and random sampling ({cyan} in Fig.~\ref{fig:xprof}). For the random sampling, 15,165 filaments were displaced on the \rosat\ maps at random positions in the same way as in T20.
The signal-to-noise ratio (S/N) can be estimated by following Eq. 4 in Sect. 3.3 of T20. The S/Ns at the filament ``cores'' (within 2 Mpc from the filament spines) result in 0.6$\sigma$(R1), 1.2$\sigma$(R2), 0.3$\sigma$(R4), 4.2$\sigma$(R5),1.9$\sigma$(R6), and 0.5$\sigma$(R7). 
We evaluated the excess of X-ray signals in two ways: allowing overlaps of different filament cores at different redshifts and not allowing the overlaps. The former provides an accumulated X-ray signal in radial directions used to estimate the overall detectability (e.g., in Fig.~\ref{fig:xprof}). 
The latter was obtained by masking cores of all the filaments, except the one under consideration, in the redshift range of 0.2<$z$<0.6. This X-ray signal gives a unique X-ray profile for the filaments and we use it for an X-ray spectral analysis to estimate physical properties of gas such as the temperature (used in Fig.~\ref{fig:xspec-fit}). After masking the overlaps of different filament cores at different redshifts, the S/Ns are -0.6$\sigma$(R1),0.6$\sigma$(R2), 0.8$\sigma$(R4), 3.1$\sigma$(R5), 1.0$\sigma$(R6), and 0.8$\sigma$(R7). The negative value of significance means that the data point has a negative value. This could happen because we evaluate the X-ray signal "after subtracting the local background".

\subsection{Gas density and temperature}
\label{subsec:temperature}

\begin{figure}
    \centering
    \includegraphics[width=\linewidth]{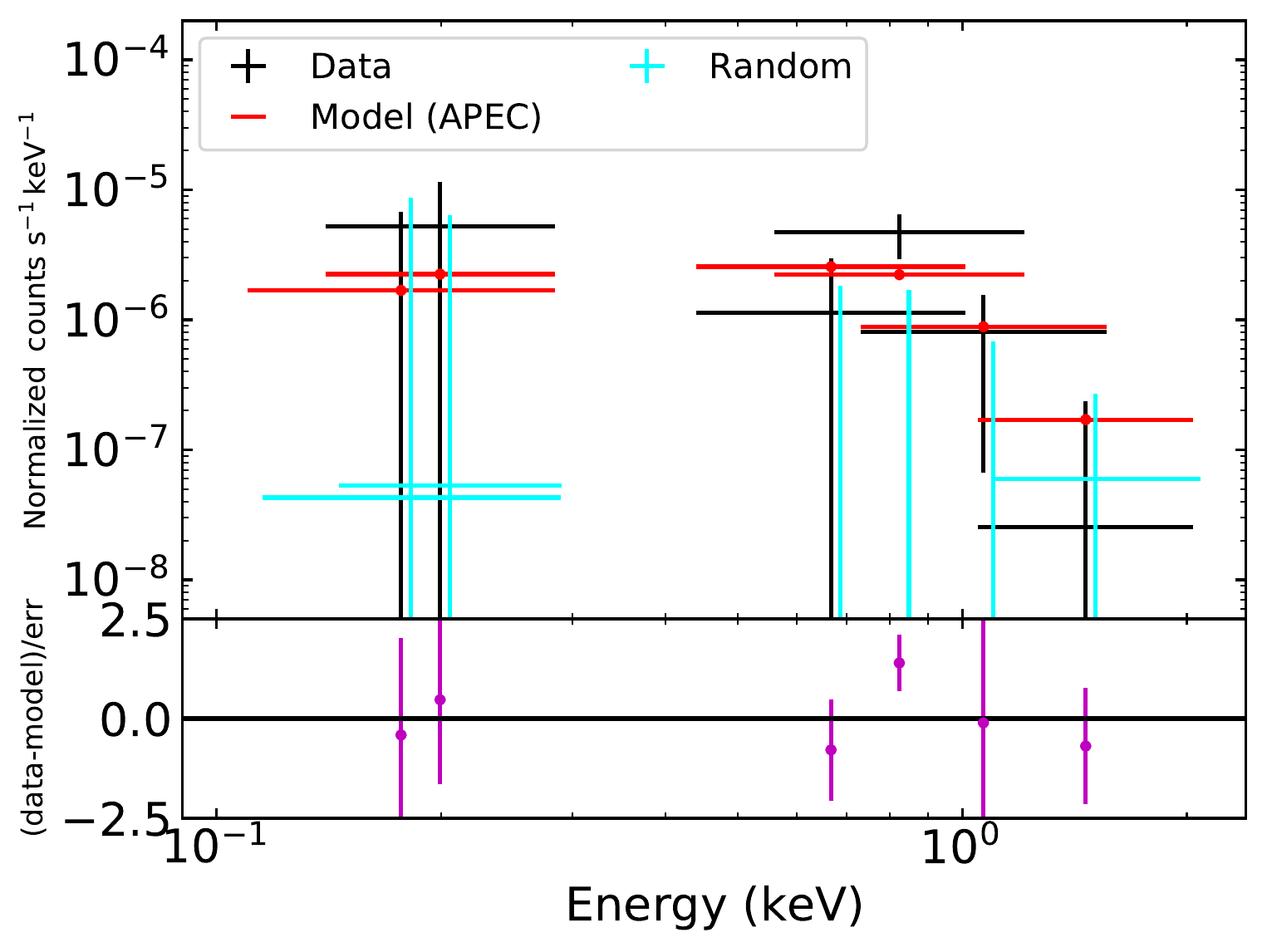}
    \caption{(Top panel) Black: Excesses of X-ray emission at the cores of 15,165 filaments (< 2 Mpc from the filament spines), relative to the background (average X-ray signal at 10--20Mpc from the filament spines), in the six \rosat\ energy bands (R1, R2, R4, R5, R6, and R7). The 1$\sigma$ uncertainties are estimated by a bootstrap resampling. Cyan: X-ray emission at the cores from 1,000 random sampling. The 1$\sigma$ uncertainties are estimated by computing a standard deviation of 1,000 random sample. The data points from the random sampling shift slightly to the right on the X-axis for a visualization. Red: The X-ray fit to the data with the APEC model (see text for details). (Bottom panel) Magenta: Ratio of data and model at each energy band. }
    \label{fig:xspec-fit}
    \end{figure}

Using the stacked X-ray count-rate measured at the six energy bands, we performed an X-ray spectral analysis using the Python X-ray spectral analysis package (PyXspec) interface with the X-ray spectral analysis package (XSPEC) \citep{Arnaud1996} to estimate the average gas density and temperature at the cores of the 15,165 filaments. 
The response file, `pspcc\_gain1\_256.rsp', is provided\footnote{https://heasarc.gsfc.nasa.gov/docs/rosat/pspc\_matrices.html}. 
For the model of the X-ray emission, we used the Astrophysical Plasma Emission Code (APEC) model \citep{Smith2001} with two free parameters: the normalization of surface brightness and temperature. The metallicity of gas was assumed to be 0.2 of the solar value \citep{anders1989}, based on the measurements in peripheries of intracluster medium (ICM) \citep{Mernier2017, Mernier2018} and predictions by hydrodynamic simulations in \cite{Martizzi2019}. The neutral atomic hydrogen (HI) column density in the region of our analysis was estimated to be $\sim2 \times 10^{20} \rm{cm^2}$ from the HI4PI map \citep{hi4pi2016}. 
We used the median redshift of our filaments, $z\sim0.44$, for the fitting.
We fit the APEC model to the data obtained by masking the overlaps of different filament cores, with a minimum chi-square method (see Fig.~\ref{fig:xspec-fit}), and find a surface brightness of $(0.06 \pm 0.03) \times 10^{-12} \, \rm erg \, cm^{-2} \, s^{-1} \, deg^{-2}$ at 0.5--2.0 keV with a gas temperature of $0.9^{+1.0}_{-0.6}$ keV. The reduced $\chi^2$ value is 0.9. 
We can calculate the overdensity of the gas in the filaments for the APEC model, using the filament model in T20 (See Section 5.1 in T20). We assumed a cylindrical filament with a density distribution following a $\beta$-model with $\beta$=2/3, including the orientation angle of the filaments on the plane of the sky. As in T20, we also assumed a negligible evolution of the overdensity in filaments in the range of our sample $0.2<z<0.6$ with a constant electron overdensity, relative to the mean electron density in the Universe. With this model and given the estimated surface brightness, the overdensity of the gas at the center of the filaments was estimated to be $30\pm15$.

Our measurements of gas density and temperature in filaments can be compared with those obtained from tSZ in T20, in which filaments in the same lengths (30--100 Mpc) and redshifts (0.2<$z$<0.6) were used. Our measured gas overdensity at the center of the filaments, $\delta=30\pm15$, is consistent with T20, who also found $\delta = 19.0^{+27.3}_{-12.1}$ at the center of the filaments  with the $\beta$ model. While our measured gas temperature, $0.9^{+1.0}_{-0.6}$ keV, is relatively higher than that of T20 (i.e., $\sim0.12 \pm 0.04$ keV), these values are consistent within 1.2$\sigma$. Considering that our gas temperature is estimated at the cores (< 2 Mpc) of the filaments whereas the value in T20 is the average within 5 Mpc, this result may imply that the gas temperature is higher at the cores compared to the outskirts. There was a hint of a radial temperature profile in a filament from hydrodynamic simulations by \cite{gheller2019}. These authors showed that the gas temperature is constant up to 2 Mpc from the filament core with a temperature of $\sim3 \times 10^6$ K ($\sim$ 0.3 keV) and starts to drop beyond the distance. However, relatively large uncertainties of our temperature measurements prevent us from drawing a definitive conclusion and more sensitive measurements would be needed to confirm it. In the following, we hence evaluate the detectability of the X-ray emission from the gas in the core of filaments we measured with a new X-ray telescope, \erosita, also considering a lower temperature gas as shown in \cite{gheller2019} and T20, in the next section.

\section{Prospects for filament detection in X-rays}
\label{sec:erosita}

As a successor of \rosat, the extended ROentgen Survey with an Imaging Telescope Array (\erosita) \citep{predehl2010} will provide a new full-sky X-ray survey thanks to its instrument aboard the Russian Spektrum-Roentgen-Gamma satellite\footnote{http://hea.iki.rssi.ru/SRG} (SRG). After completing four years of observations, \erosita\ will have made eight full sky scans, resulting in a sky-averaged exposure time of $\sim$2 ks and making \erosita\ $\sim30$ deeper than \rosat.  This will allow for the detection of about three million AGN \citep{Kolodzig2012} and about $100\,000$ galaxy clusters \citep[SRG/eROSITA Science Book:][]{Merloni2012}. In the following, we assess the detectability of X-ray emission from gas in filaments by \erosita.

\subsection{Simulating X-ray spectrum of \erosita}
\label{subsec:erosita-spec}

\begin{figure}
    \centering
    \includegraphics[width=\linewidth]{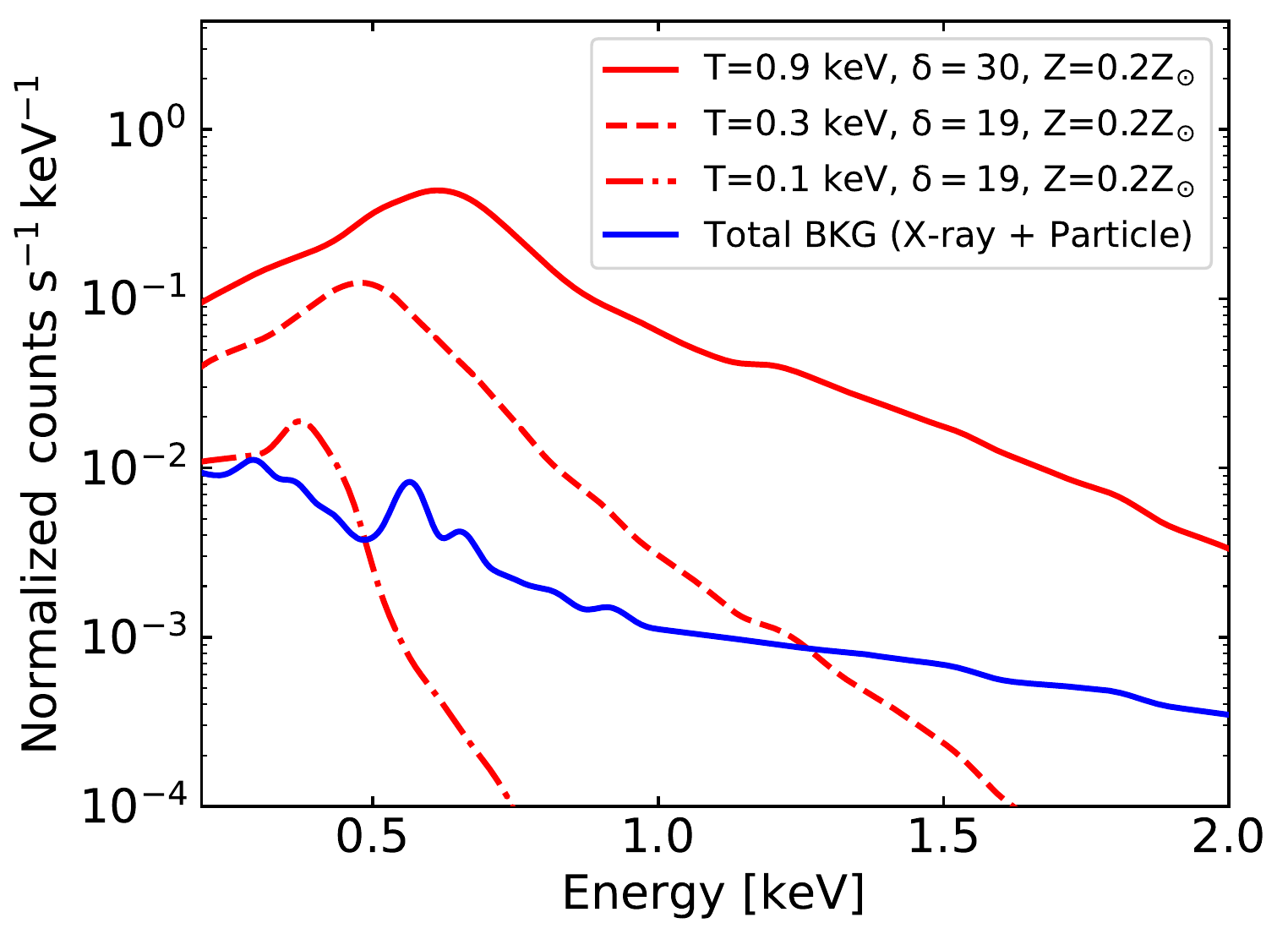}
    \caption{Red: Simulated stacked energy spectrum of \erosita\ from the gas at the cores of the stacked 15,165 filaments. The stacked spectrum is calculated with overdensity = 30, temperature = 0.9 keV, and metallicity = 0.2$\zsun$ and represented as a red solid curve. The stacked spectrum is also calculated assuming overdensity = 19, temperature = 0.3 keV and metallicity = 0.2$\zsun$ and shown as a red dashed curve, and overdensity = 19, temperature = 0.1 keV and metallicity = 0.2$\zsun$ and shown as a dash-dotted curve.  Blue: Simulated background energy spectrum of \erosita\ based on the model from \cite{Merloni2012}. }
    \label{fig:erosita-spec}
    \end{figure}

We simulated X-ray energy spectra from gas in all the 15,165 filaments as they would be observed by \erosita, including their lengths, angles, and redshifts (that determine their surface brightness on the sky and spectral shifts along the energy) via the PyXspec interface with XSPEC. We evaluated the X-ray emission at the cores of the filaments using the same model as in Sect. \ref{subsec:temperature}, with the normalization and temperature from our measured values. We also included masks of galaxy clusters, critical points from DisPerSE and X-ray points sources. We adjusted the mask size of point sources for \erosita\ to 1 arcmin in diameter \citep{Merloni2012}.  These masks reduce the effective sky areas of the filament cores to $\sim$1/3.  The simulated energy spectra of the 15,165 filaments are stacked as in Fig.~\ref{fig:erosita-spec-appendix} and the stacked spectrum is shown as a red solid curve in Fig.~\ref{fig:erosita-spec}. The background emission, shown as a blue curve, is simulated based on \cite{Merloni2012}.

We also simulated X-ray energy spectra from gas in the filaments assuming lower temperatures of 0.3 and 0.1 keV as discussed in Sect. \ref{subsec:temperature}. The normalization of the X-ray emission is calculated based on the filament model in T20, assuming a cylindrical filament with a density distribution following a $\beta$-model with $\beta$=2/3. For the amplitude of the $\beta$-model, we used the average core electron overdensity of $\delta=19$ measured in T20. We also considered inclination angles of the filaments, $\theta$, relative to the plane of the sky, as in T20. This changes the observed amplitude of X-ray emission by a factor of $1/\mathrm{cos}(\theta)$. The simulated spectra of the 15,165 filaments are stacked as in Fig.~\ref{fig:erosita-spec-appendix} and the stacked spectra are shown in Fig.~\ref{fig:erosita-spec} as red dashed and red dash-dotted curves for the gas temperatures of 0.3 keV and 0.1 keV, respectively. 

\subsection{Signal-to-noise ratio by stacking filaments}
\label{subsec:erosita-sn}

\begin{figure}
    \centering
    \includegraphics[width=\linewidth]{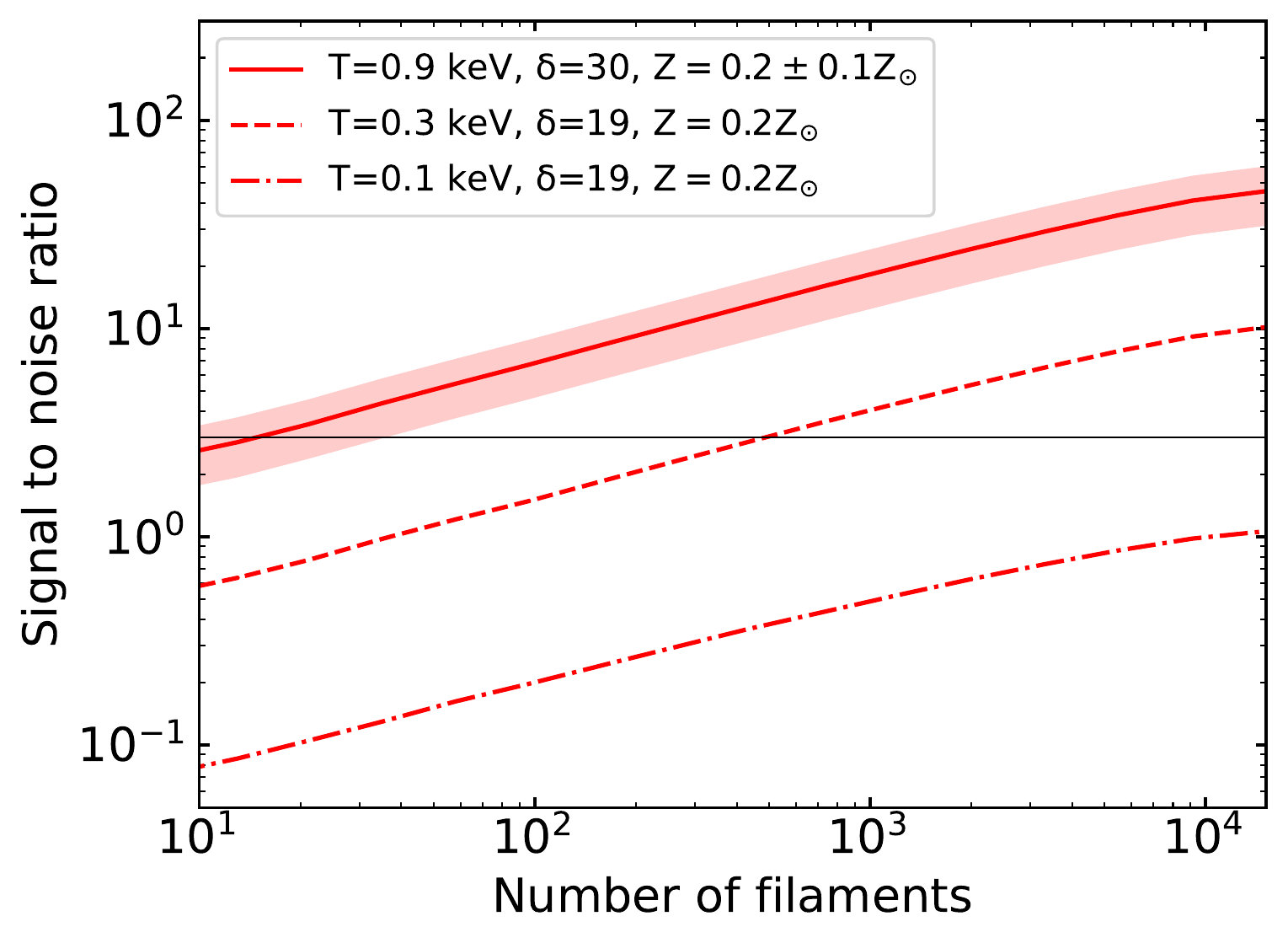}
    \caption{Signal-to-noise ratios of the X-ray emission from the gas at the cores of filaments, relative to the background, as a function of the number of stacked filaments.  Red solid curve: S/N for overdensity = 30, temperature = 0.9 keV and metallicity = 0.2$\zsun$. The S/N is evaluated at 0.5--2.0 keV to maximize it, based on Fig.~\ref{fig:erosita-spec}. Light red area: Uncertainties due to the metallicity of (0.2$\pm$0.1)$\zsun$ for overdensity = 30 and temperature = 0.9 keV. Red dashed curve: S/N for overdensity = 19, temperature = 0.3 keV and metallicity = 0.2$\zsun$, evaluated at 0.3--0.8 keV. Red dash-dotted curve: S/N for overdensity = 19, temperature = 0.1 keV and metallicity = 0.2$\zsun$, evaluated at 0.3--0.5 keV. The horizontal black line indicates a 3$\sigma$ line.}
    \label{fig:erosita-sn}
    \end{figure}

The simulated stacked spectra in Fig.~\ref{fig:erosita-spec} allow us to predict the S/Ns of the X-ray emission from the filaments relative to the background, which are summarized in Fig.~\ref{fig:erosita-sn}. The S/N is evaluated in an optimal energy range to maximize the S/N for different temperatures of gas (see Fig.~\ref{fig:erosita-spec-appendix}). The results show that we can expect a significant $\sim 46 \sigma$ detection for the gas with $T=0.9$ keV by stacking the 15,165 filaments. When the gas temperature is lower, the significance drops to $\sim 10 \sigma$ for $T=0.3$ keV (evaluated at 0.3--0.8 keV) and to $\sim 1 \sigma$ for $T=0.1$ keV (evaluated at 0.3--0.5 keV). We also vary the metallicity of gas in filaments considering 0.2$\pm$0.1$\zsun$. The result is shown as light red area for $T=0.9$ keV in Fig.~\ref{fig:erosita-sn}. We find that the effect of the metallicity is small compared to the temperature variations, but not negligible.

\section{Systematics}
\label{sec:systematics}

We investigated possible biases on the measurement of the X-ray signals from the filaments in the \rosat\ data due to resolved and unresolved galaxy groups and clusters in the foreground and background as in T20. The contribution from resolved galaxy clusters was investigated by varying the radius of the mask. We find that the resolved clusters are well covered by our mask and the contamination to our measurement is minor. In addition, the contamination from unresolved groups and clusters, which is mostly handled by the subtraction of the background signal, was investigated. The potential residual contamination is tested through the random sampling. The result of the random sampling is consistent with zero as shown in Fig.~\ref{fig:xprof}, suggesting that our results are not biased by residual contaminations from unresolved groups and clusters.

We also considered a bias from unresolved X-ray point sources such as unresolved AGN. We find that the contribution from resolved point sources is minor by comparing the X-ray profiles with and without masking all the point sources in Appendix \ref{sec:maskps}. This may imply that the contribution from unresolved point sources is also minor. In our analysis in particular, unresolved point sources at the foreground and background were subtracted by our local background subtraction since they have no positional
correlation with filaments; we added these unresolved point sources as noise. However, the contribution of point sources inside filaments could bias our result. Therefore, we tested this by performing the spectral analysis including an additional power-law model assuming that a collection of point sources in filaments has a power-law spectrum. In the test, the spectral index is varied between 1.4 and 1.7.  The result of the fitting shows that the contribution of the power-law component is only $\sim3 \times 10^{-17} \, \rm erg \, cm^{-2} \, s^{-1} \, deg^{-2}$ (0.5--2.0 keV) compared to $\sim6 \times 10^{-14} \, \rm erg \, cm^{-2} \, s^{-1} \, deg^{-2}$ (0.5--2.0 keV) of the APEC component, and we found that the contribution of point sources inside filaments was minor. To complete our analysis of the systematics related to point sources, we tested, using the Akaike's Information Criterion (AIC) method \citep{akaike1973}, whether one out of the two models (APEC and power law) can be ruled out statistically. We find that the probability that the APEC model is preferred is 64\%. This means that despite a slight preference for the APEC model, both models fit the data almost equally well. In this test, the selection of the model is so far limited by the quality of the data used; more sensitive measurements such as those of \erosita would be needed for further assessment.

We also investigated the effect of fitting all stacks with a single median redshift. To check this, we instead performed the model fitting with four APEC models with the redshifts of 0.25, 0.35 0.45, and 0.55 by linking the plasma temperatures and normalizations among them. The normalizations were linked with a weight with the relative number of filaments in each redshift bin. Other parameters were fixed. The resulting temperature and surface brightness are $\sim$0.82 keV and $\sim0.05 \times 10^{-14} \, \rm erg \, cm^{-2} \, s^{-1} \, deg^{-2}$ (0.5--2.0 keV). The values are $\sim$5\% and $\sim$10\% lower for temperature and surface brightness than the values by the fitting with a median redshift, but they are within our measurement uncertainties and not significant. 

While we assumed a baseline metallicity of 0.2$\zsun$ from the measurements in peripheries of ICM \citep{Mernier2017, Mernier2018} and predictions by hydrodynamic simulations \citep{Martizzi2019}, the metallicity in cosmic web filaments is not well known. Therefore, we repeated our analysis for 0.1$\zsun$ and 0.3$\zsun$. For the former, the average gas temperature in the filament cores is found to be $\sim 0.7$ keV and the surface brightness $\sim0.06 \times 10^{-12} \, \rm erg \, cm^{-2} \, s^{-1} \, deg^{-2}$ (0.5--2.0 keV), corresponding to an average central gas overdensity $\delta \sim 38$. If the metallicity is 0.3$\zsun$, the gas temperature is $\sim 0.9$ keV, the surface brightness is $\sim0.06 \times 10^{-12} \, \rm erg \, cm^{-2} \, s^{-1} \, deg^{-2}$ (0.5--2.0 keV), and the central gas overdensity is $\delta \sim 27$. These values are within the uncertainties of our measurement with the baseline metallicity, suggesting that no significant bias is found with the present X-ray data.

\section{Discussion}
\label{sec:discussion}

We measure for the first time, with  \rosat\ maps, the X-ray emission from the cores of large ($>30$ Mpc) cosmic filaments. 
Using \rosat\ data, \cite{Briel1995} searched for a X-ray signal from shorter filaments between pairs of galaxy clusters with a tangential distance of $\sim$20 Mpc/h on the sky. These authors accumulated the X-ray photons between the cluster pairs assuming that the filaments between the cluster pairs are almost straight, and they ended up with an upper limit that is two orders of magnitude larger than ours. 
One reason for the difference between our findings and those of \cite{Briel1995} is due to the small number of filaments stacked in that study ($\sim$40), as compared to the stack of $\sim$15,000 filaments in the present study. As shown in Fig.~\ref{fig:erosita-sn}, even in the case of \erosita\ data (30 times more sensitive than \rosat), there should be statistically enough filaments for a significant detection.
Another reason may be due to the straight-filament assumption, while those filaments may be mostly curved. This may be supported by the fact that there is no detection of filaments identified by cluster pairs with the tSZ or lensing, when the distance between clusters is larger than $\sim$15 Mpc/h, whereas several detections of shorter filaments ($\sim$10 Mpc/h) have been reported.  
It indicates the importance of identifying the precise locations of cosmic web filaments when studying them.

Our result can be compared with predictions of the extragalactic X-ray emission from hydrodynamic simulations by \cite{Phillips2001}. These authors estimated that the WHIM at $z$ = 0 constitutes about 5--15\% of the total CXB at 0.5--2 keV and that the WHIM line emission peaks at 0.5--0.8 keV. \cite{Ursino2006} also predicted that the WHIM contribution of the total CXB may reach up to 20\% in 0.37--0.925 keV, most of which is emitted by filaments at redshifts between 0.1 and 0.6. Our measurements at the \rosat\ six energy bands in Fig.~\ref{fig:xspec-fit} show an excess in the predicted energy range, which is consistent with the \asca\ and \rosat\ observations of the CXB in \cite{Miyaji1998}. 

The analysis of the stacked data in the \rosat\ six energy bands indicates that the overdensity of gas is $30\pm15$ at the centers of the filaments assuming a $\beta$-model with $\beta$=2/3. To compare to other results mostly drawn for a cylindrical filament with a uniform density distribution, we recalculated the gas overdensity from our measurements assuming a uniform density distribution of gas in cylindrical filaments within the radius of 2 Mpc (the value chosen in the present analysis). We find a value of $\delta\sim8\pm4$. In other measurements, detections of gas in individual filaments were reported with X-ray and tSZ observations in relatively short ($\sim$ 10 \mpc\ ) and dense ($\delta > 150$) filaments. Other statistical studies showed detections of lower-density gas in filaments; most of these detections were still limited to short filaments ($\sim$10 \mpc). The weak-lensing measurements of $\sim$23,000 filaments with the tangential distance of 6--10 \mpc\ in \cite{Epps2017}, the tSZ measurements of $\sim$260,000 filaments in \cite{Tanimura2019b}, and $\sim$1 million filaments in \cite{deGraaff2019} with the tangential distance of $\sim$10 \mpc\ show that the matter or gas overdensity is $\delta = 3.2\sim5.5$. A study of longer filaments with the length of 30--100 \mpc\ in T20 also estimated that the gas overdensity is $\delta \sim 6.3$, assuming a cylindrical filament with a uniform density distribution of gas in $\sim$24,000 filaments. Our measured value of gas density from the X-rays is consistent with these statistical studies of filaments.

We also estimate the gas temperature at the cores of the filaments and find $0.9^{+1.0}_{-0.6}$ keV. Our measurement can be compared with other X-ray measurements of individual filaments between cluster pairs or in the outskirts of clusters such as $\sim0.3$ keV by \suzaku\ \citep{hattori2017},  $\sim0.4$ keV by \chandra\ and \xmm\ \citep{Alvarez2018}, and 0.7 -- 2.0 keV by \xmm\ \citep{Eckert2015}.  These measurements are limited to individual, short ($\sim$ a few Mpc), and dense ($\delta$ > 200) filaments. Statistical studies were performed for short filaments ($\sim$ 10 \mpc\ )  with the tSZ measurements \citep{Tanimura2019b,deGraaff2019}. These studies found gas temperatures of $\sim 0.7$ keV and $\sim 0.3$ keV. The study of filaments of 30--100 \mpc\ length by T20 reported a temperature of $\sim0.1$ keV. Compared to these statistical studies, our measurement of gas temperature shows a relatively higher value. Considering that our gas temperature is estimated at the cores (< 2 Mpc) of the filaments, our result may imply that the gas temperature is higher at the cores, compared to the outskirts. On the other hand, the value obtained from the tSZ measurements in the \planck\ data is averaged over $\sim$5 Mpc, owing to the limited angular resolution. For a definitive statement on the gas temperature in filaments, more sensitive measurements are needed. 

\section{Conclusions}
\label{sec:conclusion}

By stacking  \rosat\ maps at the positions of $\sim$15,000 large cosmic filaments identified with the SDSS galaxies, we detect for the first time their X-ray emission at a significance of 4.2$\sigma$ in the energy band of 0.56--1.21 keV. The significance in the same band remains significant (3.1$\sigma$) even when we mask regions in which filaments at different redshifts overlap on the plane of the sky. While no significant detection is found at other energy bands, this result is predicted by hydrodynamic simulation \citep{Phillips2001,Ursino2006}, showing that the relative contribution of the WHIM to the total CXB peaks at about 0.4--0.9 keV. 

Using the stacked X-ray count rate measured at the \rosat\ six energy bands, we perform an X-ray spectral analysis with the APEC model and estimate the average gas density and temperature at the cores (< 2 Mpc) of the filaments. We find that the average central overdensity is $\delta = 30\pm15$, assuming a cylindrical filament with a density distribution following a $\beta$-model with $\beta$=2/3. We compare our measurement of gas density with other statistical measurements with the weak-lensing and the tSZ, and find that they are all consistent.
We also estimate the average gas temperature in the filament core region to be $0.9^{+1.0}_{-0.6}$ keV and find that it is a bit larger than the study of gas in large cosmic filaments with the tSZ by T20. This may imply that the gas temperature is higher at the cores compared to the outskirts, considering that the gas temperature from X-rays is estimated in a region < 2 Mpc from the filament spines whereas the values in the tSZ measurements are the average beyond the distance ($\sim$5 Mpc). 

A definitive statement on the gas temperature in filaments from the X-rays necessitates higher quality data such as those from the \erosita\ all-sky survey. As a matter of fact, when we model the emission from cosmic filaments using the gas density and temperature estimated with the \rosat\ data, stacked filaments can be detected with very high significance (i.e., $\sim 4 \sigma$ for only 100 filaments and up to $\sim 46 \sigma$ for a sample of 15,165 filaments as used here). With such high quality data as \erosita, we show that we can detect lower-temperature gas of $\sim$0.3 keV by stacking $\sim$2,000 filaments, which would lead to a $\gtrsim5\sigma$ detection

\begin{acknowledgements}
The authors thank an anonymous referee for the useful comments and suggestions. This research has been supported by the funding for the Baryon Picture of the Cosmos (ByoPiC) project from the European Research Council (ERC) under the European Union's Horizon 2020 research and innovation programme grant agreement ERC-2015-AdG 695561. The authors acknowledge fruitful discussions with the members of the ByoPiC project (https://byopic.eu/team). 
This publication has made use of the SZ-Cluster Database (\url{http://szcluster-db.ias.u-psud.fr/sitools/client-user/SZCLUSTER\_DATABASE/project-index.html}) operated by the Integrated Data and Operation Centre (IDOC) at the Institut d'Astrophysique Spatiale (IAS) under contract with CNES and CNRS. 
Maps are provided in HEALpix (http://healpix.sourceforge.net/) format \citep{gorski2005}.

\end{acknowledgements}
\bibliographystyle{aa} 
\bibliography{dfil} 


\appendix

\section{Effect of point source}
\label{sec:maskps}

\begin{figure}[ht]
    \centering
    \includegraphics[width=\linewidth]{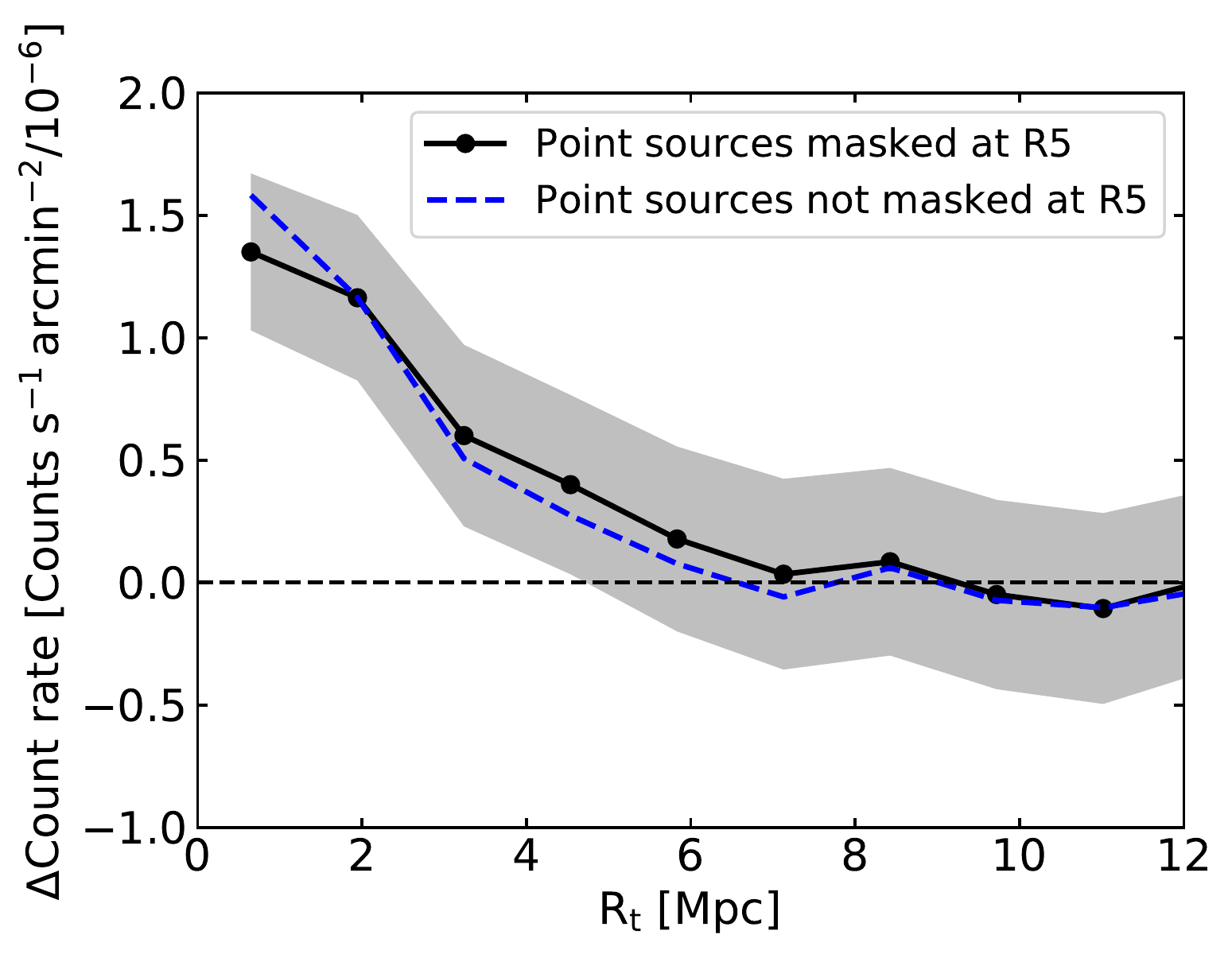}
    \caption{Average radial X-ray profiles of 15,165 filaments at R5 band with (black) and without (blue) masking resolved point sources in Sect. \ref{sec:data}, in addition to the cluster mask ($3 \times R_{500}$ + 10 arcmin). The 1$\sigma$ uncertainties of the profiles, when the point sources are masked, are estimated by bootstrap resampling and shown in gray.}
    \label{fig:xprof-noAGN}
    \end{figure}

Our measurements may be contaminated by signals from unresolved X-ray point sources. We checked the impact of resolved point sources by comparing the X-ray profiles with and without masking the resolved point sources in Fig.~\ref{fig:xprof-noAGN}; they are masked in our analysis. The resulting profiles were consistent and this suggests that the impact of the bright resolved point sources is minor. 

\section{Stacking of simulated X-ray spectra by \erosita\ }
\label{sec:erosita-hiT}

\begin{figure}
    \centering
    \includegraphics[width=\linewidth]{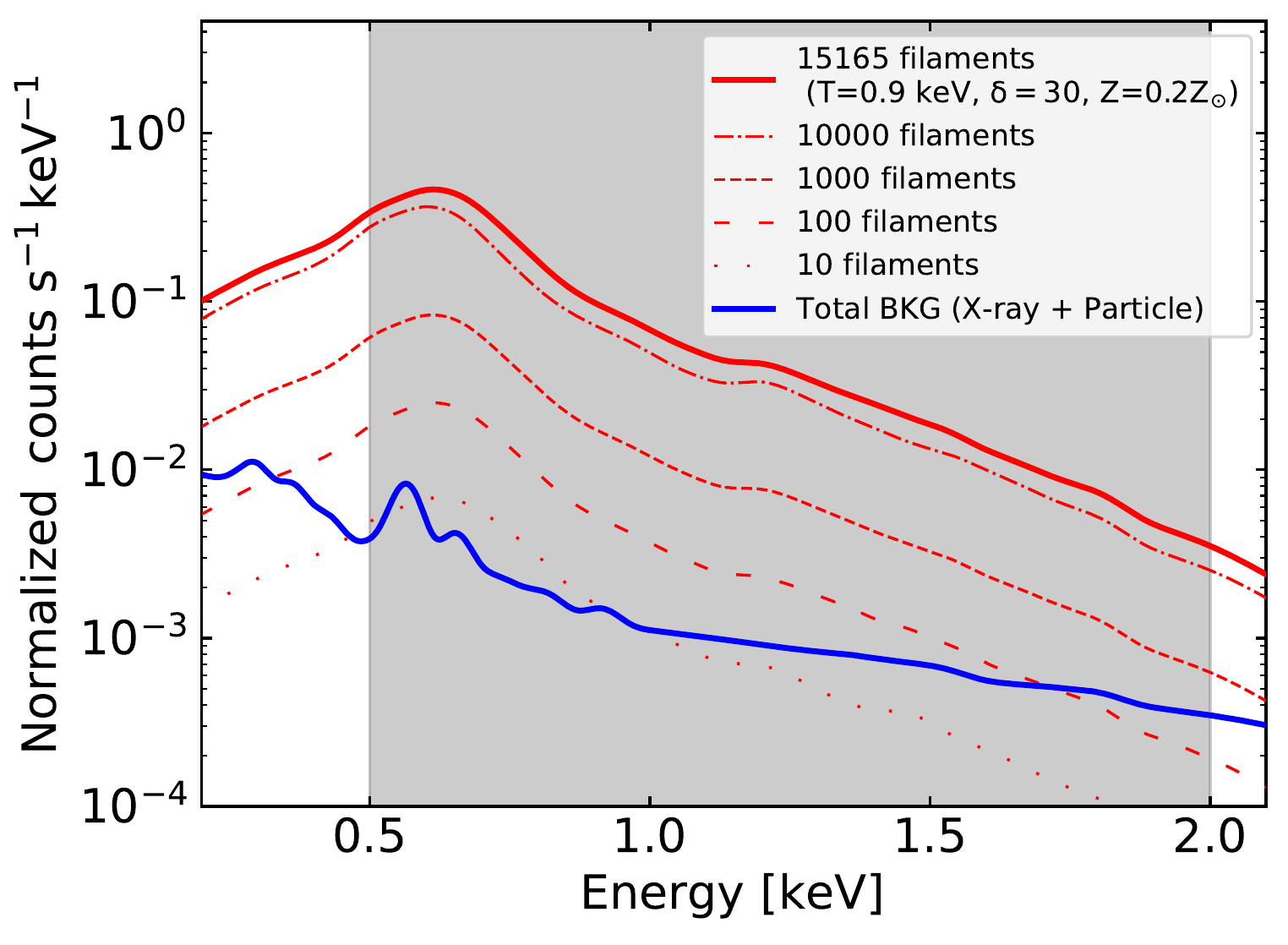}
    \includegraphics[width=\linewidth]{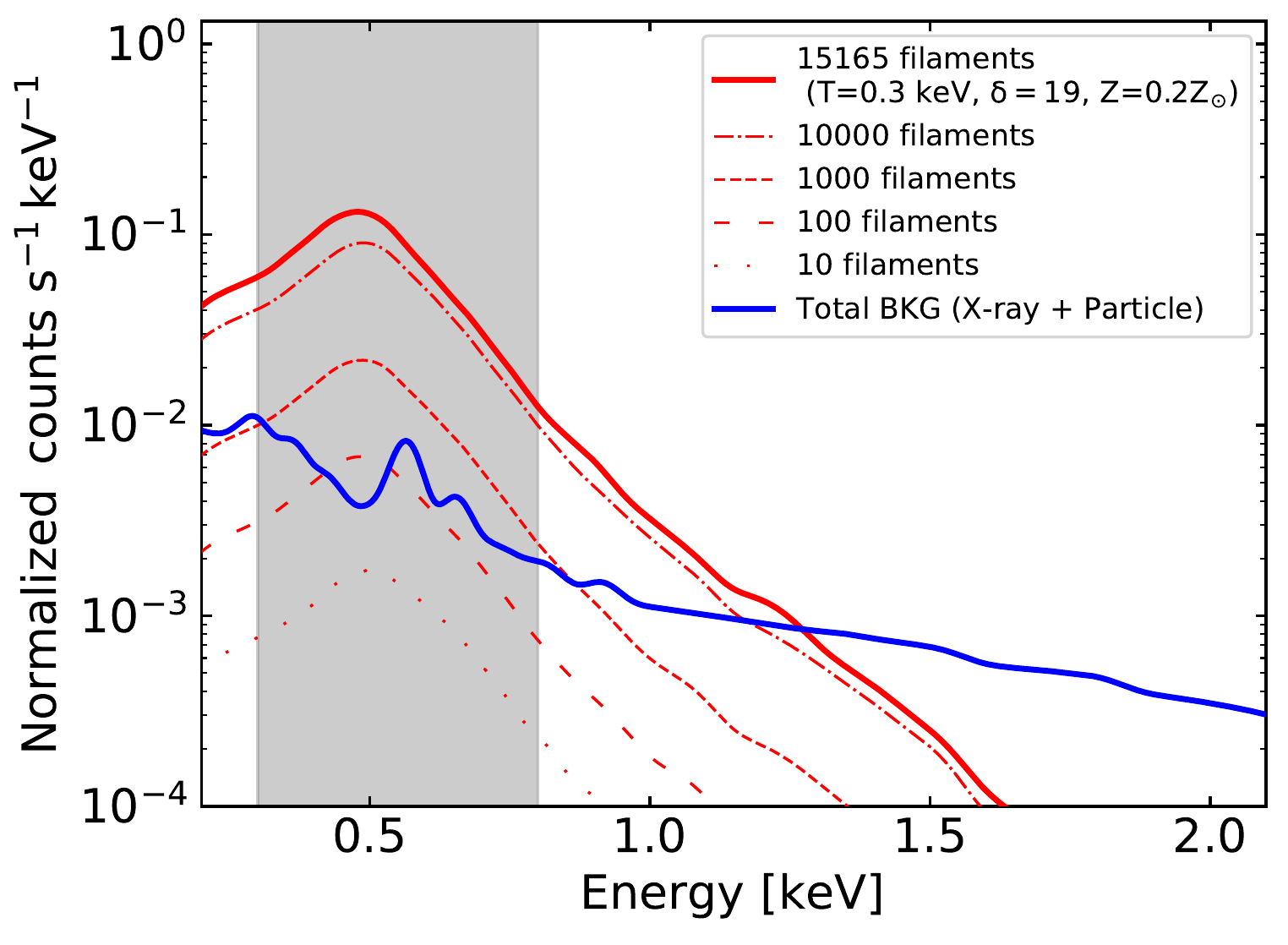}  
    \includegraphics[width=\linewidth]{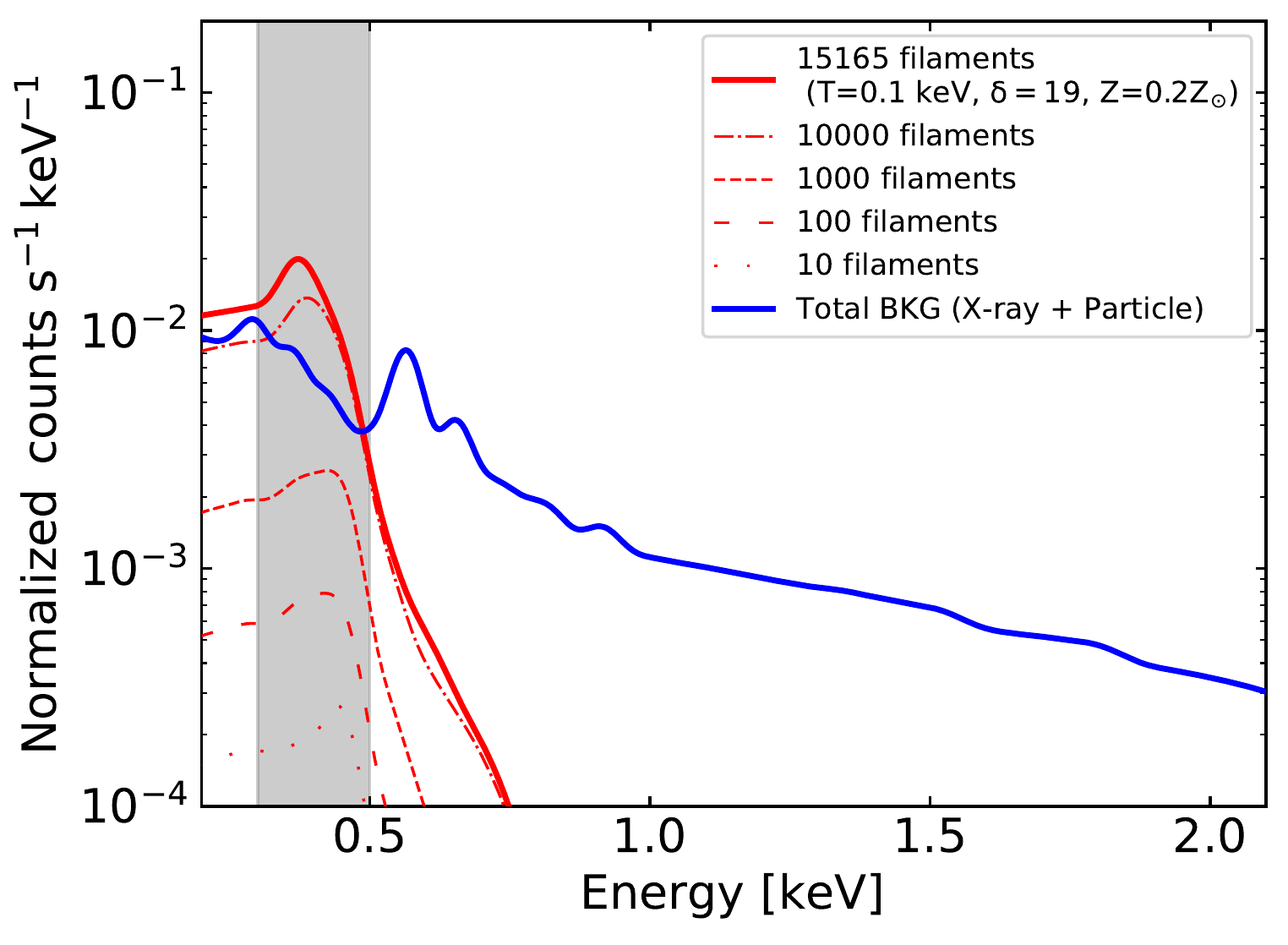}  
    \caption{Simulated energy spectra for \erosita\ from the gas at the cores of filaments. In the top panel, the spectra are calculated with an overdensity of 30, a temperature of 0.9 keV, and a metallicity of 0.2$\zsun$. The stacked spectra (red curves) display the expected signal from the brightest $N$ filaments (with $N=10$ to 15,165 ). Energy spectra are also simulated assuming overdensity of 19, temperature of 0.3 keV and metallicity of 0.2$\zsun$ in the middle panel, and overdensity of 19, temperature of 0.1 keV and metallicity of 0.2$\zsun$ in the bottom panel. Blue: Simulated background energy spectrum of \erosita. The optimal energy ranges to maximize the S/N for 15,165 filaments are shown as a light gray area.  }
    \label{fig:erosita-spec-appendix}
    \end{figure}

We simulated the X-ray energy spectra of cosmic web filaments as they would be observed by \erosita, using the APEC model described in Sect. \ref{subsec:erosita-spec}. In the top panel, the spectra were calculated with an overdensity of 30, temperature of 0.9 keV, and metallicity 0.2$\zsun$. The stacked spectra (red curves in Fig. \ref{fig:erosita-spec-appendix}) display the expected signal from the brightest $N$ filaments (with $N=10$ to 15,165). In the middle panel (bottom panel), the spectra were computed with an overdensity of 19, a temperature of 0.3 (0.1) keV with a metallicity 0.2$\zsun$. 
The background emission, shown as a blue curve, was simulated based on  \cite{Merloni2012}. For each case, we identified the optimal energy range to maximize the S/N for 15,165 filaments, and we found that it is 0.5--2.0 keV for the gas temperature of 0.9 keV, 0.3--0.8 keV for 0.3 keV, and 0.3--0.5 keV for 0.1 keV, respectively; these are shown as a light gray area.

\end{document}